\documentclass{article}
\usepackage{spconf,amsmath,graphicx,hyperref}

\usepackage{comment}
\usepackage{cite}
\usepackage{amssymb,amsfonts}
\usepackage{algorithmic}
\usepackage{textcomp}
\usepackage{xcolor}

\usepackage{subcaption}
\usepackage{adjustbox}
\usepackage{multirow}
\usepackage{arydshln}
\usepackage{url}
\usepackage{rotating}
\usepackage{algorithm}
\usepackage{algorithmic}
\usepackage{textcomp}
\usepackage{bbm}
\usepackage{makecell}
\title{ASR-Synchronized Speaker-Role Diarization}
\name{Arindam Ghosh, Mark Fuhs, Bongjun Kim, Anurag Chowdhury, Monika Woszczyna}
\address{Solventum Health Information Systems, USA}
\begin{document}
\ninept

\maketitle

\begin{abstract}
Speaker-role diarization (RD), such as doctor vs. patient or lawyer vs. client, is practically often more useful than conventional speaker diarization (SD), which assigns only generic labels (speaker-1, speaker-2). The state-of-the-art end-to-end ASR+RD approach uses a single transducer that serializes word and role predictions (role at the end of a speaker's turn), but at the cost of degraded ASR performance. To address this, we adapt a recent joint ASR+SD framework to ASR+RD by freezing the ASR transducer and training an auxiliary RD transducer in parallel to assign a role to each ASR-predicted word. For this, we first show that SD and RD are fundamentally different tasks, exhibiting different dependencies on acoustic and linguistic information. Motivated by this, we propose (1) task-specific predictor networks and (2) using higher-layer ASR encoder features as input to the RD encoder. Additionally, we replace the blank-shared RNNT loss by cross-entropy loss along the 1-best forced-alignment path to further improve performance while reducing computational and memory requirements during RD training. Experiments on a public and a private dataset of doctor–patient conversations demonstrate that our method outperforms the best baseline with relative reductions of $6.2\%$ and $4.5\%$ in role-based word diarization error rate (R-WDER), respectively.
\end{abstract}

% without degrading ASR performance.
% 6.5 to 6.1 (6.2% relative) on DoPaCo and 2.2 to 2.1 (4.5% relative) on SiMeCo.


\vspace{-0.2cm}
\section{Introduction}
\vspace{-0.2cm}

% Anurag comments:
% 1. more ground information -- conversations, need to differentiate speaker, diarize who spoke what. 
% 3. RD is a different problem.
% 4. why SD? why RD? why is it important?
% why we need single model for ASR and RD
% What are the challenges of the problem? This is the motivation of doing this work.
% Our contribution is based on this motivation.

% Pattern of the Introduction section:
% What is the topic/problem and why is it important to study it, what is the use of it? -> What are the research issues/challenges with this problem (this is the motivation for everyone to solve it)? -> What others have done till now to solve it? -> What are the weakness/drawback/limitation of their approach? (This is my motivation to propose a method and solve it) -> My contribution (what all we did? How much it provided improvement? What did we find - analysis?)

% talk about lack of role-based data in the introduction (challenges).
% Use the different pattern for ASR 2nd
% Discussion about when there is no audio, what happens to the role predicttion.
% Double quotation
% Limitation and Future work
% Is rnnt loss, memory intensive or compute intensive?
% 3. There is no dataset in the literature to study the problem.
% [need to write somewhere about why we did not use a public dataset for role diaraization - because of the lack of it.]
% Check all the bold fonts in the notations
% correct the (t,u)-steps

\begin{figure*}[t]
    \centering
    \begin{subfigure}[t]{0.18\textwidth}
    \centering
        \includegraphics[scale=0.28]{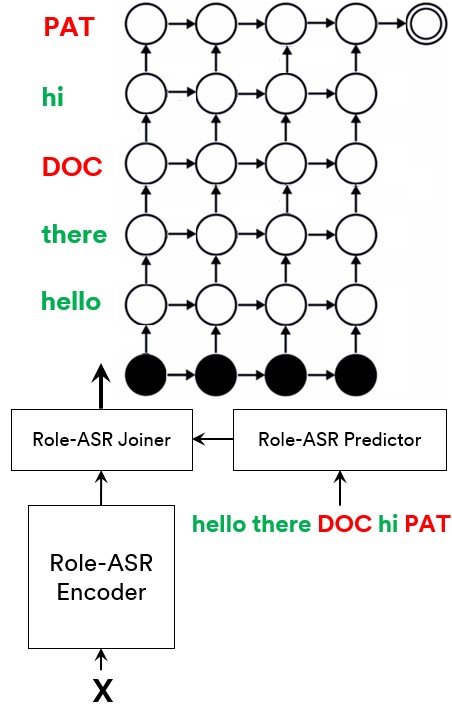}
        \caption{Role-ASR \cite{shafey19_interspeech}.}
        \label{subfig:role_asr}
    \end{subfigure}
    ~\vrule\vrule~
    \begin{subfigure}[t]{0.38\textwidth}
    \centering
        \includegraphics[width=\textwidth]{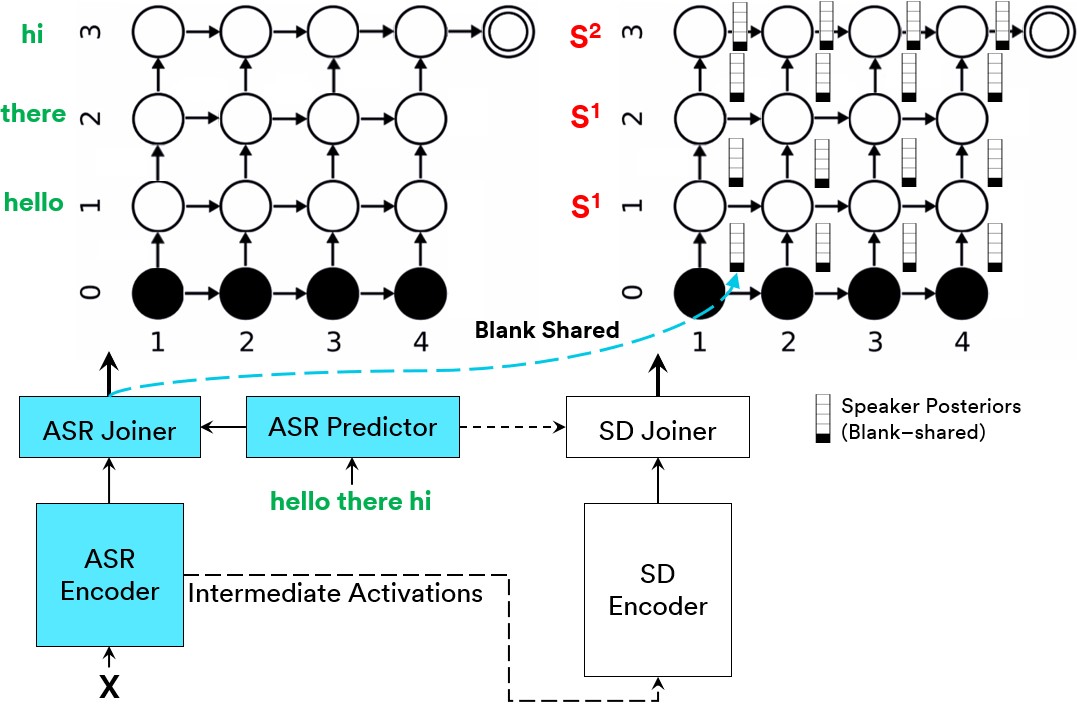}
        \caption{ASR-Synchronized SD \cite{huang24d_interspeech, raj24_odyssey}.}
        \label{subfig:blank_shared_hat_sa_surt}
    \end{subfigure}
    ~\vrule\vrule~
    \begin{subfigure}[t]{0.39\textwidth}
    \centering
        \includegraphics[width=\textwidth]{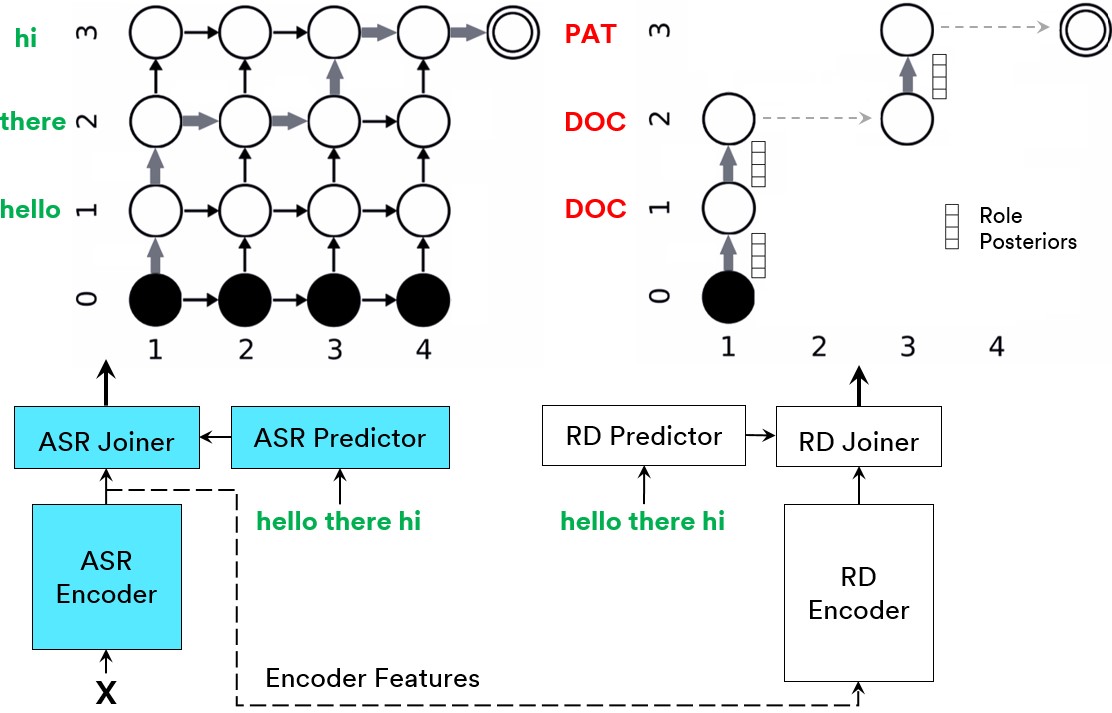}
        \caption{ASR-Synchronized RD.}
        \label{subfig:our_system}
    \end{subfigure}
    
    \vspace{-0.1cm}
    \caption{Illustration of prior work in (a) and (b), along with our proposed RD training in (c), for an example audio $\mathbf{x}$ with corresponding ground-truth label sequences: $\mathbf{y}^{\text{Role-ASR}} = [\text{hello}, \text{there}, \text{DOC}, \text{hi}, \text{PAT}]$, $\mathbf{y}^{\text{ASR}} = [\text{hello},\text{there},\text{hi}]$, $\mathbf{y}^{\text{SD}} = [\text{s}^1,\text{s}^1,\text{s}^2]$, and $\mathbf{y}^{\text{RD}} = [\text{DOC},\text{DOC},\text{PAT}]$. Blue colored ASR modules remain frozen during the training of auxiliary SD and RD transducers. The graphs at the top of each network shows the alignment paths used to compute the training loss. In (c), the path with bold arrows represent the 1-best forced-alignment path.}
    \label{fig:asr_rd_systems}
    \vspace{-0.3cm}
\end{figure*}

% \textbf{What is the paper/problem about? - Joint ASR and Speaker Diarization}
In multi-speaker conversations like doctor-patient encounters, lawyer-client interactions, meetings, etc., ASR captures ``what was spoken" while speaker diarization (SD) determines ``who spoke when." Combining the two, joint ASR + SD determines ``who spoke what," which improves the intelligibility of, otherwise lengthy and unstructured, transcripts and helps downstream natural language processing (NLP) tasks, such as summarization, information extraction, etc., with more accurate and meaningful insights.

% \textbf{What is RD? Why is it used?}
An issue with SD is that it assigns generic labels (e.g. speaker-1, speaker-2) without any prior knowledge of their identities \cite{landini2022_vbx}. For real-world applications, such as summarization, identifying the \textit{speaker-role} (doctor, patient, lawyer, client, etc.) is more valuable than knowing the actual speaker identity. Although one can utilize external speaker profiles to map the generic SD labels to actual speakers, enabling role inference, such external data is often unavailable. In these scenarios, joint ASR + speaker-role diarization (RD) that directly determines ``which role spoke what" becomes crucial.

% \textbf{What are the challenges of RD, and what has been done in the literature?}
An approach to RD is to infer roles directly from ASR transcripts \cite{zuluagagomez2023_slt, wang2024_diarizationlm, wu2025_text_sd}. For this, large language models (LLMs) have also been shown to be effective to some extent \cite{wang2024_diarizationlm, wu2025_text_sd}. However, text-only approaches struggle when transcripts lack distinguishing linguistic cues, for example around short back-and-forth utterances (“yeah,” “right”), due to speakers using similar vocabulary, or due to ASR errors. In these cases, acoustic information becomes essential for reliable separation.

% \textbf{What has been done on using speech+text for RD in the literature? Modular systems, unified systems?}
Existing approaches combining acoustic and linguistic information for speaker-role diarization \cite{valente2011_icassp, sapru2014_icassp, flemotomos2020_odyssey, flemotomos2022_interspeech, blatt2024_interspeech} fall into two categories: modular and end-to-end models. Modular systems, such as \cite{flemotomos2020_odyssey}, use role-oriented language models for classification, refined by x-vector clustering. In \cite{flemotomos2022_interspeech}, a BERT-based classifier is integrated with role-constrained x-vector clustering. However, these systems suffer from error propagation, which makes end-to-end models more appealing. Early works in this area employed a transducer model to perform ASR + RD by inserting role tokens at speaker turns in the training text \cite{shafey19_interspeech}, and explored a unified ASR + RD model for air traffic control conversations \cite{blatt2024_interspeech}.

% \textbf{WER-drop problem in ASR+RD systems. How to solve the issue of WER drop?}
Integrating ASR and RD into a single model poses its own challenges. Direct multitask learning leads to ASR performance degradation \cite{shafey19_interspeech, kim25k_interspeech}, a concern also noted for joint ASR+SD \cite{chang22_interspeech, huang24d_interspeech}. To mitigate this, \cite{chang22_interspeech} trained the ASR and the turn-prediction transducers separately to preserve the word error rate (WER). For joint ASR + SD, \cite{huang24d_interspeech} addressed this by freezing the ASR transducer and training an auxiliary SD transducer in sync via blank logit sharing (ASR-synchronized SD), a technique also adopted in \cite{raj24_odyssey} for speaker-attributed ASR.

Building on the success of ASR-synchronized SD \cite{huang24d_interspeech}, in this paper, we propose an ASR-synchronized RD model that predicts a role per every ASR-predicted word in parallel. For this, we show that speaker diarization and role diarization are fundamentally different tasks, requiring different acoustic and linguistic contexts, and propose appropriate modifications to the network architecture and the training strategy to tailor it for role prediction.
% Additionally, we propose an RD-guided blank-suppression heuristic to reduce deletion errors in ASR decoding.
The main contributions of this work are as follows.
\begin{enumerate}
    \item We provide, to the best of our knowledge, the first analysis of the relative importance of acoustic and linguistic modalities for joint ASR+RD. While SD relies primarily on acoustic differences, the contribution of these modalities for RD, which involves simultaneous speaker differentiation and role inference, is not directly clear. Our results establish that RD relies more heavily on linguistic context than on acoustic features.

    \item Based on this finding, we modify the baseline ASR-sync. SD model \cite{huang24d_interspeech} for joint ASR+RD by explicitly injecting linguistic information into the RD network. This is achieved through (1) task-specific predictors (CNN for ASR and RNN for RD) and (2) the use of higher-layer ASR encoder features as input to the RD encoder. These modifications substantially improve R-WDER, with relative reductions of $19.23\%$ on a private dataset (DoPaCo) and $54.5\%$ on a public dataset (SiMeCo).
    
    % from $7.8 \rightarrow 6.3$ on a private dataset (DoPaCo) and $5.5 \rightarrow 2.5$ on a public dataset (SiMeCo).

    \item We further propose a simplified training strategy for the auxiliary RD transducer by replacing the blank-shared RNNT loss \cite{huang24d_interspeech} with a simpler cross-entropy loss computed along the 1-best forced-alignment path. This not only reduces computational and memory requirements during training, but also yields additional relative R-WDER reductions of $3.2\%$ on DoPaCo and $16\%$ on SiMeCo.

    % from $6.3 \rightarrow 6.1$ on DoPaCo and $2.5 \rightarrow 2.1$ on SiMeCo.

\end{enumerate}

\section{Preliminaries}
In this section, we revisit the Role-ASR \cite{shafey19_interspeech} and ASR-synchronized SD \cite{huang24d_interspeech} models that serve as baselines for this paper. Let $\phi$ be the blank token, $\{w^1,\ldots, w^{N_{\textbf{ASR}}}\}$ denote the set of subwords, $\{\text{s}^1,\ldots,\text{s}^{N_{\text{SD}}}\}$ the set of speakers, and $\{\text{r}^1,\ldots,\text{r}^{N_{\text{RD}}}\}$ the set of speaker-roles, where $N_{\text{RD}} \leq N_{\text{SD}}$. Let a transducer model be denoted by $M$, where $M \in \{\text{Role-ASR}, \text{ASR}, \text{SD}, \text{RD}\}$ as shown in Figure \ref{fig:asr_rd_systems}. For a given audio $\mathbf{x} =[x_1, \ldots, x_{T}]$ and the ground-truth label sequence $\mathbf{y} = [y_1, \ldots, y_{U}]$, let the $M$-encoder's output be $[f^M_1(\mathbf{x}), \ldots, f^M_{T}(\mathbf{x})]$ and the $M$-predictor's output be $[g^M_1, \ldots, g^M_{U}]$, where $f^M_t \in \mathbb{R}^{N_f}$ and $g^M_u \in \mathbb{R}^{N_g}$, with $t$ and $u$ denoting the frame and label indices and $N_f$ and $N_g$ representing the encoder and predictor dimensions, respectively. Then, $M$-joiner's raw logits $j^{\text{M}}_{t,u}$ are given by
\begin{align}
    h^{\text{M}}_{t,u} &= P^{\text{M}}\cdot f^{\text{M}}_t(x_{1:T}) + Q^{\text{M}} \cdot g^{\text{M}}_u(y_{1:u-1}) + b^{\text{M}}_h \label{eq:h_tu} \\
    j^{\text{M}}_{t,u} &= A^{\text{M}} \cdot \tanh (h^{\text{M}}_{t,u}) + b^{\text{M}}_s \quad\quad \in \mathbb{R}^{|\mathcal{V}_{\text{M}}|} \label{eq:s_tu}
\end{align}
where $P^{\text{M}}$, $Q^{\text{M}}$ and $A^{\text{M}}$ are projection matrices, $b^{\text{M}}_h$ and $b^{\text{M}}_s$ are bias terms, and $|\mathcal{V}^{\text{M}}|$ is the vocabulary size. Let $\mathcal{A}(\mathbf{y})$ be the set of all possible alignments between $\mathbf{y}$ and $\mathbf{x}$, more specifically between $\mathbf{y}$ and $f^{M}_{1:T}(\mathbf{x})$.

\subsection{Baseline 1: Role-ASR \cite{shafey19_interspeech}}
\label{sec:role_asr}
RNNT model \cite{graves2012_arxiv} was originally proposed for ASR as an improvement over the CTC model. Shafey et al. \cite{shafey19_interspeech} first used it for joint ASR and RD by inserting role tokens at the end of each speaker's turn, e.g. $\mathbf{y} = [\text{hello}, \text{there}, \text{DOC}, \text{hi}, \text{PAT}]$, and training directly using the RNNT loss, as shown in Figure \ref{subfig:role_asr}. For this model, using Equation \ref{eq:h_tu} and \ref{eq:s_tu}, the probability distribution over the vocabulary $\hat{y}_{t,u} \in \mathcal{V}^{\text{Role-ASR}} = \{w^1,\ldots, w^{N_{\textbf{ASR}}}\} \cup \{\text{r}^1,\ldots,\text{r}^{N_{\text{RD}}}\} \cup \{\phi\}$ is
\begin{align*}
    &\mathbb{P}(\hat{y}_{t,u} | x_{1:T}, y_{1:u-1}) = \text{softmax}(j^{\text{Role-ASR}}_{t,u}),
\end{align*}
and the RNNT loss for label $\mathbf{y}$ above is given by
\begin{align}
    &\mathcal{L}_{\text{RNNT}} = -\ln \mathbb{P}(\mathbf{y}|x_{1:T}) = -\ln \sum_{a \in \mathcal{A}({\mathbf{y}}) } \mathbb{P}(a | x_{1:T}) \nonumber \\
    &= -\ln \hspace{-0.2cm} \sum_{a \in \mathcal{A}({\mathbf{y}}) } \prod_{\hat{y}_{t,u} \in a} \mathbb{P}(\hat{y}_{t,u} | x_{1:T}, y_{1:u-1}). \label{eq:rnnt_loss}
\end{align}

We use this model with an RNN predictor as our first baseline system B1: Role-ASR (RNN) in Table \ref{tab:total_wer_wder_dopaco_eval}.

\subsection{Baseline 2: ASR-Synchronized SD \cite{huang24d_interspeech}}
\label{sec:asr_sync_sd}
% The framework in \cite{huang24d_interspeech} proposes an ASR-synchronized SD model that predicts a word and a corresponding speaker label in parallel. As illustrated in Figure \ref{subfig:blank_shared_hat_sa_surt}, the model uses an auxiliary SD transducer for per-word speaker prediction, which is synchronized with an already trained ASR transducer through blank sharing. The SD transducer receives input from an intermediate layer of the ASR encoder and uses the ASR predictor network. The ASR transducer is frozen during SD training.

Huang et al. \cite{huang24d_interspeech} introduced an ASR+SD model that jointly predicts words and corresponding speaker labels using two synchronized hybrid auto-regressive transducers (HAT) \cite{variani2020_icassp}, as shown in Figure \ref{subfig:blank_shared_hat_sa_surt}. First, the ASR transducer is trained independently using the vocabulary $\mathcal{V}^{\text{ASR}} = \{w^1,\ldots, w^{N_{\textbf{ASR}}}\} \cup \{\phi\}$. Using Equation \ref{eq:h_tu} and \ref{eq:s_tu}, the probabilities of blank and non-blanks, $\hat{y}_{t,u} \in \mathcal{V}^{\text{ASR}}$, are given by
{\small
\begin{align}
    &P(\hat{y}_{t,u} =\phi | x_{1:T}, y_{1:u-1}) = \text{sigmoid}(j^{\text{ASR}}_{t,u}[0]) =: b_{t,u} \label{eq:b_tu}\\
    % &P(\hat{y}_{t,u} = w^{1:{N_{\textbf{ASR}}}}| x_{1:T}, y_{1:u-1}) \nonumber \\
    % &\hspace{.5cm}= (1 - b_{t,u}) \cdot \text{softmax}(j^{\text{ASR}}_{t,u}[1:{N_{\textbf{ASR}}}]). \nonumber
    &P(\hat{y}_{t,u} = w^{i}| x_{1:T}, y_{1:u-1}) = (1 - b_{t,u}) \cdot \text{softmax}(j^{\text{ASR}}_{t,u}[i]), \nonumber
\end{align}
}%
for $i \in \{1, \ldots, N_{\textbf{ASR}}\}$. Using these, the RNNT loss for an example ASR sequence $\mathbf{y} = [\text{hello},\text{there},\text{hi}]$ is obtained from Equation \ref{eq:rnnt_loss}.

Next, the ASR transducer is frozen and the SD transducer is trained using blank-shared RNNT loss (BS-RNNT) with the vocabulary $\mathcal{V^{\text{SD}}} \cup \{\phi\}$, where $\mathcal{V^{\text{SD}}} = \{\text{s}^1,\ldots,\text{s}^{N_{\text{SD}}}\}$ and $\phi$ is borrowed from ASR. The SD encoder takes features from an intermediate (5th) layer of the ASR encoder as input. Since SD predictor, a 1-D convolutional layer with a kernel size of $2$ (CNN-$2$), is also borrowed from the ASR transducer, $g^{\text{SD}}_u = g^{\text{ASR}}_u$. While the blank posterior is the same as in Equation \ref{eq:b_tu}, the probability of speaker labels, $\hat{y}_{t,u} \in \mathcal{V}^{\text{SD}}$, are given by
\begin{align*}
    P(\hat{y}_{t,u} | x_{1:T}, y_{1:u-1}) = (1-b_{t,u})\cdot \text{softmax}(j^{\text{SD}}_{t,u}).
\end{align*}
Using these, the RNNT loss for a corresponding speaker-label sequence $\mathbf{y} = \{\text{s}^1,\text{s}^1,\text{s}^2\}$ is obtained from Equation \ref{eq:rnnt_loss}.

% \begin{equation*}
%     \mathcal{L}_{\text{SD}}
%     = -\ln \hspace{-0.2cm} \sum_{a \in \mathcal{A}_{\mathbf{s}}} \prod_{s_{t,u} \in a} \mathbb{P}(s_{t,u} | f^{\text{SD}}_{1:T}, g^{\text{ASR}}_{1:u-1}).
% \end{equation*}

During inference, whenever ASR emits $\hat{y}_{t,u} \neq \phi$, the corresponding speaker label is predicted by applying an \(\text{argmax}(\cdot)\) over the SD logits $j^{\text{SD}}_{t,u}$. We apply this model as it is to RD to obtain our second baseline B2: ASR (CNN-$2$) + RD (CNN-$2$, L5) BS-RNNT in Table \ref{tab:total_wer_wder_dopaco_eval}.

\section{Proposed ASR-Synchronized RD}
\label{sec:asr_sync_rd}
In this section, we propose modifications to the ASR-Synchronized SD model in terms of (1) predictor network, (2) ASR encoder's features as input to RD encoder, and (3) training criterion to make it suitable for ASR-Synchronized RD. The rationales for these proposals are as follows.

\subsection{Role Diarization Relies More on Linguistic Features}
\label{sec:pred_context_matters}

% \vspace{-0.2cm}
\begin{figure}[ht]
    \centering
    \begin{subfigure}[t]{0.48\linewidth}
    \centering
        \includegraphics[width=\linewidth]{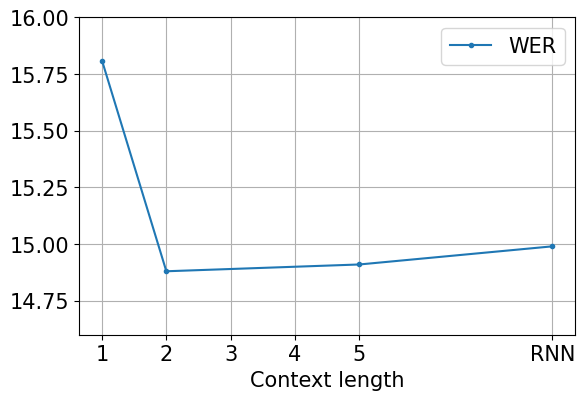}
        \vspace{-0.5cm}
        \caption{Word-prediction capability.}
    \end{subfigure}
    ~
    \begin{subfigure}[t]{0.46\linewidth}
    \centering
        \includegraphics[width=\linewidth]{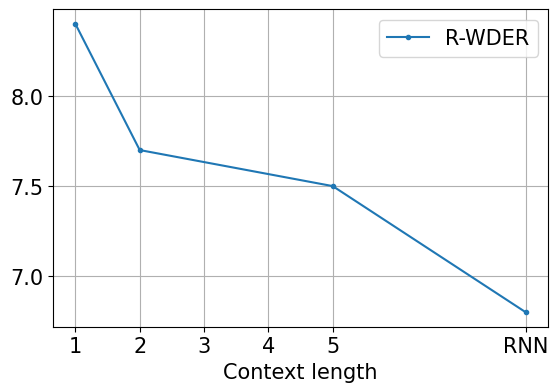}
        \vspace{-0.5cm}
        \caption{Role-prediction capability.}
    \end{subfigure}
    \vspace{-0.2cm}
    \caption{Effect of Role-ASR predictor's context length on (a) WER and (b) R-WDER when trained on DoPaCo and evaluated on the validation set of DoPaCo.}
    \label{fig:role_asr_context_length}
    \vspace{-0.2cm}
\end{figure}

% To claim it's an LM, you'd need to show that it is modeling future word likelihood. There are plenty of NLP classification tasks that benefit from long-range dependency modeling that aren't LMs, per se. Perhaps the stronger argument to make here is that ASR and role classification are different tasks with different architectural demands, so separating them proves beneficial.

Intuitively, SD is expected to rely primarily on speaker-specific acoustic cues rather than linguistic context. In contrast, RD involves not only differentiating between speakers, but also determining the roles they play in the interaction, which necessitates both acoustic and semantic (i.e., what was spoken) information. The relative importance of these two modalities, however, is not immediately clear for RD. In the following, through two complementary studies, we demonstrate that RD is more strongly influenced by linguistic context than by acoustic features.

One source of linguistic context in a transducer model is the predictor network. Using the Role-ASR model (Section \ref{sec:role_asr}), we show that while ASR may benefit from a smaller predictor's context \cite{variani2020_icassp, ghodsi2020_icassp, prabhavalkar2021_icassp}, role prediction definitely requires a longer context. For this, in Figure \ref{fig:role_asr_context_length}, we plot the word-prediction capability (WER) and the role-prediction capability (R-WDER defined in Section \ref{subsec:evaluation}) of the Role-ASR model versus the predictor's context length. To model a finite context length of $n$, we use CNN-$n$, and for an infinite context, we use an RNN. We observe that while WER remains mostly unchanged (CNN-$2$ being the best), R-WDER improves significantly as we increase the context length. This confirms that, unlike word prediction, role prediction benefits from the long-range linguistic information provided by the predictor. Therefore, the proposed system P1 differs from baseline system B2 by using separate task-specific predictors: CNN-$2$ for ASR and RNN for RD. We label this system as P1: ASR (CNN-$2$) + RD (RNN, L5) BS-RNNT in Table \ref{tab:total_wer_wder_dopaco_eval}.

The second source of linguistic information is the word-level features from the upper layers of the ASR encoder. We show that while speaker prediction works best with features from intermediate layers (5th layer in \cite{huang24d_interspeech}) that encode more speaker-level information, role prediction works best with features from upper layers that store more word-level information. For this, we take system P1 from above and use it with features from different layers of the ASR encoder. In Table \ref{tab:rwder_layer}, we see that R-WDER improves as we go from lower to higher layers for both the DoPaCo and SiMeCo sets. Note that since the model has been trained only on DoPaCo, R-WDER on out-of-domain SiMeCo is comparatively much worse; however, the trend is the same. Based on this, we modify the system P1 to instead use the last layer (L12) features and denote it as P2: ASR (CNN-$2$) + RD (RNN, L12) BS-RNNT in Table \ref{tab:total_wer_wder_dopaco_eval}.

\begin{table}[htb]
    \normalsize
    \centering
    \begin{adjustbox}{max width=\linewidth}
    \begin{tabular}{r|cc}
        \makecell{ASR Enc. Layer} &DoPaCo Val &SiMeCo Val \\
        \cline{1-3}
        L-2 &8.1 &42.1 \\
        L-5 &7.2 &40.3 \\
        L-8 &\textbf{6.9} &34.3 \\
        L-10 &7.0 &30.9 \\
        L-12 &\textbf{6.9} &\textbf{29.5} \\
    \end{tabular}
    \end{adjustbox}
    \caption{R-WDER performance of the model ASR (CNN-2) + RD (RNN, L-$n$) BS-RNNT, with RD encoder taking input from different layers (L-$n$) of the ASR encoder. The model is trained on DoPaCo and evaluated on DoPaCo and out-of-domain SiMeCo.}
    \label{tab:rwder_layer}
    \vspace{-0.3cm}
\end{table}

% Note that, for training the ASR (CNN-$2$), we use the vanilla RNNT loss instead of the blank-factorized RNNT loss used in \cite{huang24d_interspeech} as we found vanilla RNNT to produce better WER. For the blank-shared baselines, we simply derive the blank probability from the ASR logits and pass it to the RD transducer which then uses it for blank-factorized RNNT loss along with the role probabilities from the RD transducer.

% Next, we propose the final modification to the original baseline \cite{huang24d_interspeech} in its loss function that further improves the role prediction performance.

\vspace{-0.2cm}
\subsection{1-Best Alignment Path + Cross Entropy Loss}
\label{sec:fa_ce}
The baseline system B2 \cite{huang24d_interspeech} employs blank-factorized HAT transducers \cite{variani2020_icassp} so that the ASR transducer's blank-emission probability, at each $(t,u)$ step, can be shared with the auxiliary transducer, providing a kind of soft alignment information along all RNNT paths (Figure \ref{subfig:blank_shared_hat_sa_surt}). The auxiliary transducer then uses the full RNNT loss (computation over all alignment paths) to train its corresponding emissions.

For our proposed method, we consider only the 1-best alignment path and train the RD-transducer simply using cross-entropy loss. For this, as shown in Figure \ref{subfig:our_system}, the reference text is first forced aligned with the ASR-transducer to identify the $(t,u)$ steps at which non-blank tokens are emitted. Then, using the cross-entropy loss, the RD network is trained to emit role tokens at its corresponding $(t,u)$ steps. For example, for the ASR label sequence $\mathbf{y} = [\text{hello},\text{there},\text{hi}]$, let $ \mathbf{a^{*}} = [\text{hello}, \text{there}, \phi, \phi, \text{hi}, \phi] \in \mathcal{A}(\mathbf{y})$ be the 1-best forced-aligned path, and $ \mathbf{a^{*}_{r}} = [\text{DOC}, \text{DOC}, -, -, \text{PAT}, -]$ be the corresponding role label path (Figure \ref{subfig:our_system}), where $-$ indicates no training on blank emissions. Then, with the vocabulary
$\mathcal{V^{\text{RD}}} = \{\text{r}^1,\ldots,\text{r}^{N_{\text{RD}}}\}$, the role probabilities, for $\text{r}_{t,u} \in \mathcal{V}^{\text{RD}}$, and the cross-entropy-based RD loss are given, respectively, by
\begin{align}
    &P(\text{r}_{t,u} | x_{1:T}, y_{1:u-1}) = \text{softmax}(j^{\text{RD}}_{t,u}), \nonumber \\
    &\mathcal{L_{\text{RD}}} \hspace{0.1cm} = \hspace{-0.3cm} \sum_{r_{t,u} \in \mathbf{a^{*}_{r}},~r_{t,u} \neq -} \hspace{-0.5cm} -\ln \mathbb{P}(r_{t,u} | x_{1:T}, y_{1:u-1}). \nonumber \label{rd_loss_noblank}
    %&\mathcal{L_{\text{RD}}} = \sum_{r_{t,u} \in \mathbf{a^{*}_{\textbf{r}}} } \hspace{-0.3cm} -\mathbbm{1}(r_{t,u} \neq -) \cdot \ln \mathbb{P}(r_{t,u} | f^{\text{RD}}_{1:T}, g^{\text{RD}}_{1:u-1}), \label{rd_loss_noblank}
\end{align}
During inference, similar to \cite{huang24d_interspeech}, at $(t,u)$ steps where ASR emit non-blank tokens, we apply $\text{argmax}(\cdot)$ on the RD logits $j^{\text{RD}}_{tspeaker ,u}$ to predict the corresponding speaker role. 

Using this training method, our final proposed model P3: ASR (CNN-2) + RD (RNN, L12) 1Best-CE, in Table \ref{tab:total_wer_wder_dopaco_eval}, outperforms the system trained with blank-shared (soft) alignment + RNNT loss. Unlike the RNNT loss that is computed over multiple possible alignment sequences, the 1-best alignment provides a deterministic target sequence, simplifying the optimization landscape and reducing the computational and memory overhead associated with marginalizing over the alignment matrix. This enables larger batch sizes and overall shorter training time. Moreover, while RNNT loss must be recomputed for all training samples at every epoch, forced alignments can be computed once and reused across subsequent epochs, further improving efficiency.

% The former also eliminates the need for a blank-factorized HAT transducer, whose WER on the DoPaCo dataset ($16.68$) is worse than that of a vanilla transducer ($15.67$).

% Lastly, the main argument against the use of forced alignment, in the literature, has been the requirement of a pre-trained ASR model; however, note that this requirement is already part of our problem setup: Given a pre-trained ASR, how do we train an auxiliary model for synchronous role prediction?

\vspace{-0.3cm}
\section{Experiments}

\begin{table*}[t]
    \normalsize
    \caption{WER and R-WDER performance of different baseline and proposed models trained on DoPaCo and evaluated on DoPaCo and out-of-domain SiMeCo. The column with ``finetuned" refers to DoPaCo-trained models further finetuned on SiMeCo and evaluated on SiMeCo. The type of the predictor and the ASR encoder layer from which RD encoder takes features as input are shown in parentheses. The best results are shown in bold. Since systems B2, P1, P2, and P3 share the same frozen ASR module, the WER is reported only once.}
    \vspace{-0.2cm}
    \centering
    \begin{adjustbox}{max width=\linewidth}
    \begin{tabular}{c|l|l|cc|cc|cc}
        &   &   &\multicolumn{2}{c|}{DoPaCo Eval} &\multicolumn{2}{c|}{SiMeCo Eval} &\multicolumn{2}{c}{SiMeCo Eval (finetuned)} \\
        \cline{4-9}
        &Model &Model Size &WER &RWDER &WER &RWDER &WER &RWDER \\
        \cline{1-9}
        
        \multirow{3}{*}{\rotatebox[origin=c]{90}{Baseline}}
        &B1: Role-ASR (RNN) \cite{shafey19_interspeech} &104M &16.03 &6.5 &10.62 &34.2 &\textbf{8.8} &2.2 \\
        &ASR (CNN-$2$) + &62.4M + &\textbf{15.67} &- &\textbf{10.56} &- &\textbf{8.8} &-\\
        &B2:\quad RD (CNN-$2$, L5) BS-RNNT Loss \cite{huang24d_interspeech} &~~41.6M=104M &'' &7.8 &'' &44.3 &'' &5.5 \\
        \cline{1-9}
        
        \multirow{3}{*}{\rotatebox[origin=c]{90}{Ours}}
        &P1:\quad RD (RNN, L5) BS-RNNT Loss &~~42.6M=105M &'' &7.1 &'' &41.3 &'' &5.0 \\
        &P2:\quad RD (RNN, L12) BS-RNNT Loss &~~42.6M=105M &'' &6.3 &'' &28.9 &'' &2.5 \\
        &P3:\quad RD (RNN, L12) 1Best-CE Loss &~~42.6M=105M &'' &\textbf{6.1} &'' &\textbf{27.2} &'' &\textbf{2.1}
        
        % \cdashline{2-9}
        % & & & & & & & & \\
        % \cdashline{2-9}
        % &\quad\quad~ P3 with 6 blocks &~~29.6M=92M & &6.4 & &28.7 & & \\
        % &\quad\quad~ P3 with 3 blocks &~~16.6M=79M & &6.6 & &28.5 & & \\
    \end{tabular}
    \end{adjustbox}
    \label{tab:total_wer_wder_dopaco_eval}
    \vspace{-0.35cm}
\end{table*}

%  Same as the main (above) table in the paper, but with WDER inluded.
% \begin{table*}[t]
%     \normalsize
%     \centering
%     \begin{adjustbox}{max width=\linewidth}
%     \begin{tabular}{c|l|l|cc|cc|cc}
%         &   &   &\multicolumn{2}{c|}{DoPaCo Eval} &\multicolumn{2}{c|}{SiMeCo Eval} &\multicolumn{2}{c}{SiMeCo Eval (finetuned)} \\
%         \cline{4-9}
%         &Model &Model Size &WER &RWDER (WDER) &WER &RWDER (WDER) &WER &RWDER (WDER) \\
%         \cline{1-9}
        
%         \multirow{3}{*}{\rotatebox[origin=c]{90}{Baseline}}
%         &B1: Role-ASR (RNN) &104M &16.03 &6.5 (6.1) &10.62 &34.2 (31.8) &\textbf{8.8} &2.2 (2.2) \\
%         &ASR (CNN-$2$) + &62.4M + &\textbf{15.67} &- &\textbf{10.56} &- &\textbf{8.8} &-\\
%         &B2:\quad RD (CNN-$2$, L5) BS-RNNT Loss &~~41.6M=104M &'' &7.8 (6.5) &'' &44.3 (37.3) &'' &5.5 (5.5) \\
%         \cline{1-9}
        
%         \multirow{3}{*}{\rotatebox[origin=c]{90}{Ours}}
%         &P1:\quad RD (RNN, L5) BS-RNNT Loss &~~42.6M=105M &'' &7.1 (6.0) &'' &41.3 (36.2) &'' &5.0 (5.0) \\
%         &P2:\quad RD (RNN, L12) BS-RNNT Loss &~~42.6M=105M &'' &6.3 (6.0) &'' &28.9 (27.5) &'' &2.5 (2.5) \\
%         &P3:\quad RD (RNN, L12) 1Best-CE Loss &~~42.6M=105M &'' &\textbf{6.1} (5.8) &'' &\textbf{27.2} (25.8) &'' &\textbf{2.1} (2.1)
%     \end{tabular}
%     \end{adjustbox}
%     \label{tab:wer_rwder_wder}
% \end{table*}

\vspace{-0.2cm}
\subsection{Dataset}
\label{subsec:dataset}
For our experiments, we use an internal dataset of real-world doctor-patient conversations (DoPaCo) and a public dataset of simulated medical conversations (SiMeCo) \cite{fareez2022_simeco}. 

DoPaCo consists of conversations recorded with microphones placed at medium-to-far-field positions within medical exam rooms, covering a wide range of medical specialties. These are manually transcribed and anonymized, where speakers in a conversation are labeled according to their functional role: doctor, patient, other1, other2, etc. The dataset is divided into training (1750h), validation (26.5h), and test (35.0h) sets. 

SiMeCo consists of simulated near-field conversations between only two roles: doctor and patient. Of the total 272 recordings, we remove 9 that appear to be telephony. The remaining 263 conversations are from the specialties of respiratory (207), musculoskeletal (46), cardiology (5) and gastroenterology (5). Each specialty is split in a 60:20:20 ratio to form training (21.3h), validation (6.75h), and evaluation (8.5h) sets, respectively. 

On all recordings, we apply a TDNN-LSTM-based voice activity detector to split them into segments of length typically less than 18 sec. For text tokenization, we use SentencePiece \cite{kudo2018_sentencepiece}, in unigram mode \cite{kudo2018_subword}, with a vocabulary of size 5000.

\vspace{-0.2cm}
\subsection{Model Architecture}
The ASR encoder contains two convolutional subsampling layers that reduce the frame rate by a factor of 4, followed by 12 e-branchformer layers \cite{kim23_slt}, each with a dimensionality of 384, 6-headed self-attention, and feedforward projection sublayers with 1536 dimensions. The RD encoder is similar to ASR encoder but with 9 e-branchformer layers and convolutional layers replaced by linear layers to preserve the frame rate. The RNN predictor is a single-layer LSTM with 384 dimensions, and the CNN-$2$ predictor is a 1-D convolutional layer with a kernel size of 2. Role-ASR transducer is made up of a 21-layer e-branchformer encoder and a 2-layer RNN predictor to match the size of the dual-transducer systems. The joint network of all transducers has 512 hidden dimensions.

% We initialize the RD predictor with a compatible pre-trained model from Role-ASR: the CNN-$2$ RD predictor is initialized from the Role-ASR (CNN-$2$) predictor, while the RNN RD predictor is initialized from Role-ASR (RNN) predictor.

\vspace{-0.2cm}
\subsection{Training}

Models are trained using the ESPnet framework \cite{watanabe2018_espnet} using up to 20 sec long audio segments processed into 64-dim log-Mel features with speed perturbation \cite{ko2015_interspeech} and SpecAugment \cite{park19_interspeech} applied.
%We use the ESPnet framework \cite{watanabe2018_espnet} for model training. We use 64-dim log-Mel features extracted over a window of 25ms with a frame shift of 10ms. For data augmentation, we apply speed perturbation \cite{ko2015_interspeech} and SpecAugment \cite{park19_interspeech}. The maximum length of the input audio is set to 20 seconds.

We train the Role-ASR and ASR transducers on DoPaCo for 40 epochs using Adam optimizer with an initial learning rate of $2\times10^{-3}$, warm-up scheduler with $15$k warm-up steps, and a weight decay of $10^{-6}$. For SiMeCo, because of its small size, we initialize the Role-ASR and ASR transducers with their corresponding DoPaCo-trained weights and then fine-tune them for 30 epochs with an initial learning rate of $10^{-4}$, warm-up scheduler with $10$k warm-up steps, and a weight decay of $10^{-6}$. We average the ten best models based on the validation loss to get the final model.

For both DoPaCo and SiMeCo, we train the RD transducer from scratch for 30 epochs, with an initial learning rate of $10^{-4}$, warm-up scheduler with $1$k warm-up steps, and a weight decay of $10^{-6}$. We average the ten best models based on the validation R-WDER to get the final model.

% For the encoders of Role-ASR and ASR, we apply an additional CTC loss with a weight of $0.3$.

% Dopaco-pretrained ASR and Role-ASR models have good performance on Simeco, at least in terms of WER which otherwise will be difficult to leran from scratch from such a small amount of data, whereas the RD models have very poor R-WDER, thus they are not good source of initialization.

% The train and validation set of both datasets single-speaker utterances. Thus, we apply multi-seg augmentation to both the sets, wherein we merge some of the segments (max gap <1.5s and max length < 20s) to create utterances that include multiple speakers speaking in turn.

\vspace{-0.1cm}
\subsection{Inference and Scoring}
\label{subsec:evaluation}
For ASR and Role-ASR, we use beam search \cite{graves2012_arxiv} with a beam of size 20. For RD, we follow the $\text{argmax}$ method in Section \ref{sec:asr_sync_rd}. 

For scoring, we use \emph{asclite} from the NIST Scoring Toolkit (SCTK) \cite{nist2021_asclite} that accounts for multi-speaker alignments. The ASR performance is measured using WER and role prediction performance is measured using role-based WDER (R-WDER). R-WDER is computed similarly to WDER \cite{shafey19_interspeech} but words hypothesized to be spoken by the doctor can only be scored as correct against reference doctor words; same for the patient. Other-role words can match words from any single other reference speaker.

\vspace{-0.2cm}
\subsection{Results}
\label{sec:results}

In Table \ref{tab:total_wer_wder_dopaco_eval}, we have the baseline models B1 and B2 as described in Section \ref{sec:role_asr} and Section \ref{sec:asr_sync_sd}, respectively. When trained on DoPaCo and evaluated on in-domain DoPaCo, we see that ASR (CNN-$2$) achieves the better WER ($15.67$ vs. $16.03$) while Role-ASR (RNN) achieves the better R-WDER ($6.5$ vs. $7.8$). When evaluated on out-of-domain SiMeCo, we observe the effect of domain mismatch with both models performing poorly, especially in terms of R-WDER ($34.2$ and $44.3$). When these DoPaCo-trained models are further finetuned on SiMeCo, we see the expected improvements, with both achieving the same WER ($8.8$) and Role-ASR again outperforming in terms of R-WDER ($2.2$ vs. $5.5$).

% We see that, while CNN-$2$ predictor is better at word prediction than RNN predictor (for both ASR and Role-ASR models), RNN is better at role prediction. This again confirms our finding that ASR and RD require different predictor context lengths for optimal performance. 
% We first compare the performance of the RD loss with (Equation \ref{rd_loss_blank}) and without (Equation \ref{rd_loss_noblank}) the blank token. For this, we use the ASR+RD model with common frozen CNN-$2$ predictor, denoted by ASR (CNN-$2$) + RD.

The reason for the poor R-WDER performance of baseline B2, as discussed in Section \ref{sec:pred_context_matters}, is the lack of injection of the necessary linguistic information into the RD network. Thus, applying the proposed modifications of (1) task-specific RNN predictor and (2) use of ASR encoder's 12th-layer features, we get systems P1 and P2, respectively, that achieve significant R-WDER improvements across all settings ($7.8$\textrightarrow$6.3$, $44.3$\textrightarrow$28.9$, $5.5$\textrightarrow$2.5$). Finally, applying the 1-best alignment + cross-entropy loss, proposed in Section \ref{sec:fa_ce}, we get our final system P3 that achieves the best R-WDER in all cases.

Overall, the results show that the joint ASR+RD model, with separate task-specific predictors and trained only on the 1-best force-aligned ASR path using cross-entropy loss, outperforms the baseline models in terms of both WER and R-WDER.

\vspace{-0.1cm}
\section{Conclusion}
This work demonstrated that a modified version of the auxiliary speaker diarization network \cite{huang24d_interspeech} can be successfully adapted for speaker-role diarization, achieving superior performance over the integrated approach in \cite{shafey19_interspeech}. The study also established that, unlike speaker prediction, role prediction depends more on linguistic context than on acoustic cues, motivating two modifications: (1) separate task-specific predictors: CNN for ASR and an RNN for role prediction, and (2) the use of higher-layer ASR encoder features which encode richer linguistic information. In addition, replacing the RNNT loss with a cross-entropy loss along the 1-best forced-alignment path simplified the training of the role prediction network and significantly improved performance. Future work will explore training strategies leveraging the middle ground of $n$-best paths.

\vfill\pagebreak

% References should be produced using the bibtex program from suitable BiBTeX files (here: strings, refs, manuals). The IEEEbib.bst bibliography style file from IEEE produces unsorted bibliography list.

\bibliographystyle{IEEEbib}
\bibliography{references}

\newpage
\appendix
\onecolumn
\noindent {\Large \textbf{Appendix}}

\section{Speaker-Role Diarization Guided ASR Decoding}

In this section, we show how feedback from auxiliary RD transducer can be used to improve ASR decoding. Unlike \cite{huang24d_interspeech, raj24_odyssey}, where the ASR transducer is untouched to avoid interference from the auxiliary task, we use the RD posteriors to propose a blank-suppression heuristic that can reduce deletion errors.

As shown in Table \ref{tab:wer_sub_del_ins}, the ASR transducer with a CNN-$2$ predictor experiences high deletion errors on DoPaCo. A potential solution to recover these deletions, without modifying the model, is to suppress the incorrect blanks and promote the deleted correct word tokens during the beam search. However, the challenge remains: how do we identify the $(t,u)$-steps in the beam search where blank suppression is needed?

\begin{figure*}[h]
    \centering
    \includegraphics[width=\linewidth, height=7cm]{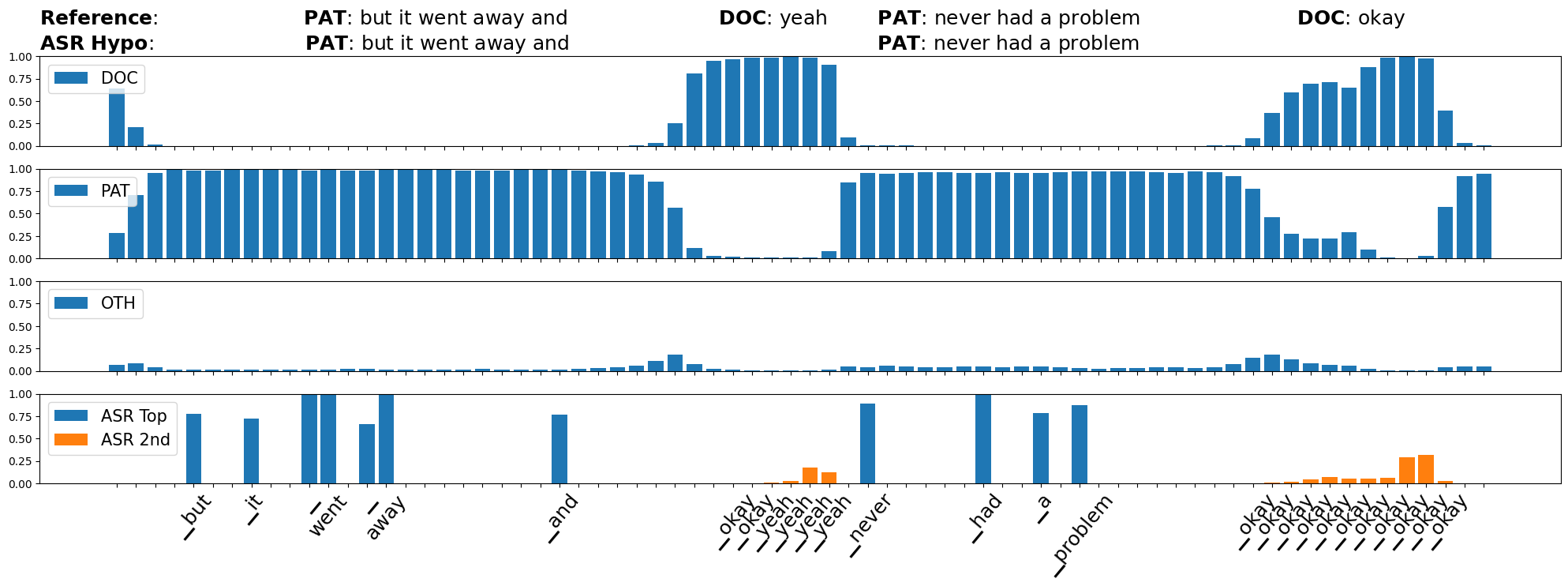}
    \vspace{-0.6cm}
    \caption{Activity of ASR and RD networks at each $(t,u)$-step of the best beam search path. The reference and hypothesis are shown at the top. The top three plots show the RD network's posteriors for DOC, PAT, and OTH. The bottom-most plot shows the ASR posteriors for the top token in blue (blank token omitted to avoid clutter) and the second-best token (for two specific regions) in orange.
    }
    \label{fig:posterior_activity}
    \vspace{-0.2cm}
\end{figure*}

For this, in Figure \ref{fig:posterior_activity}, we plot the posterior activity of our ASR (CNN-$2$) + RD (RNN) model for the utterance ``PAT: but it went away and; DOC: yeah; PAT: never had a problem; DOC: okay". The bottom-most plot shows that ASR deletes ``yeah" and ``okay" spoken by DOC, as blank is the most likely token in those regions. The top three plots reveal that although the RD network is trained only on frames where a subword token is emitted, its activity at other frames indicates an awareness of speaker activity. Notably, the RD network correctly detects DOC’s activity (top-most plot) in the deleted regions. If we plot the second-best ASR tokens (orange in the bottom-most plot), we find that they are indeed ``yeah" and ``okay", suggesting that RD activity can provide valuable signals for blank suppression.

Based on this, we propose a blank-suppression heuristic, outlined in Algorithm \ref{alg:blank_suppression}, wherein we first identify the set of top-$n$ deleted tokens $D_n$ from the validation set. Then, at any $(t,u)$-step of the ASR beam search, where the top non-blank token has a sufficiently high posterior ($\ge \alpha$) and belongs to $D_n$, and the RD network confidently ($\geq \beta$) detects role activity, we transfer the probability mass from blank to non-blank tokens. For DoPaCo, we tune the parameters using the validation set, setting $\alpha = 0.1$, $\beta = 0.99$, and $D_n = \{\text{yeah}, \text{okay}\}$. Additionally, to prevent excessive insertions, we enforce a minimum gap of three steps between consecutive suppressions.

\begin{algorithm}[h]
\begin{algorithmic}[1]
    \STATE Set $\alpha$, $\beta$, $D_n$ (set of top-$n$ deleted tokens from val set).
    \FOR{every $(t,u)$-step along a beam search path}
        \STATE $p^{\text{ASR}}_{t,u} = \text{softmax}(j^{\text{ASR}}_{t,u-1})$
        \STATE $p^{\text{RD}}_{t,u} = \text{softmax}(j^{\text{RD}}_{t,u-1})$
        \IF{$\text{argmax}(p^{\text{ASR}}_{t,u,1:{N_{\textbf{ASR}}}}) \in D_n ~\land~ \max(p^{\text{ASR}}_{t,u,1:{N_{\textbf{ASR}}}}) \ge \alpha ~\land~ \max(p^{\text{RD}}_{t,u}) \ge \beta $}
            \STATE $p^{\text{ASR}}_{t,u,\phi} = 0.01$ \hspace{0.3cm} \# Set it to some low value
            \STATE $p^{\text{ASR}}_{t,u} = p^{\text{ASR}}_{t,u} / \sum_{v} p^{\text{ASR}}_{t,u,v}$
        \ENDIF
    \ENDFOR
\end{algorithmic}
\caption{RD-Guided ASR Blank Suppression}
\label{alg:blank_suppression}
\end{algorithm}

\begin{figure}[h]
    \centering
    \includegraphics[width=0.4\linewidth]{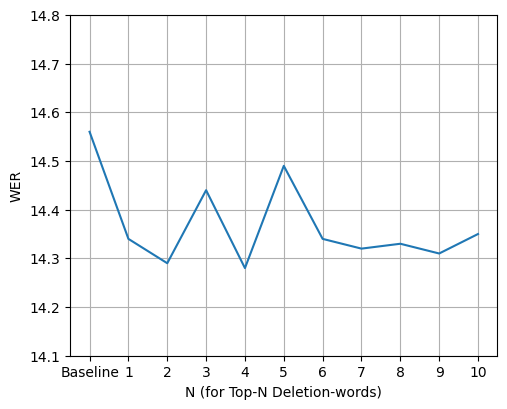} 
    \caption{DoPaCo validation set: reduction in WER when top-$n$ ($n \in [1,2,\ldots,10$) deletion-words are accounted for during RD-guided ASR decoding. The top-$10$ deletion-words for the validation set are: yeah, okay, I, and, it, you, a, the, oh, right.}
    \label{fig:correction_top_n_deletions}
\end{figure}

We find that applying the heuristic to all tokens does not yield any overall improvement because while this helps to reduce deletions for some tokens, it also increases the insertions. Too much rise in insertions will eventually nullify the improvement due to reduction in deletions. Thus, we first use the val set to identify the tokens that have the most deletions and then selectively apply our blank-suppression to them. In order to do so, in Figure \ref{fig:correction_top_n_deletions}, we plot the WER versus $n \in \{1,\ldots, 10\}$, where top-$n$ deletion tokens are applied blank suppression. 
We find that the most benefit we get from RD-guided decoding is when we apply it only top-$2$ deletion tokens.

In Table \ref{tab:wer_sub_del_ins}, we see that the ASR (CNN-$2$) + RD (RNN) model has disproportionately high deletions compared to insertions. To reduce them, we apply our proposed blank suppression heuristic in Algorithm \ref{alg:blank_suppression} and see that it improves the WER of the frozen ASR on both val and eval sets ($14.56$\textrightarrow$14.29$, $15.67$\textrightarrow$15.65$). From Table \ref{tab:wer_sub_del_ins}, we further see that RD-guided decoding is successful in doing its intended job of reducing deletion errors ($6.47$\textrightarrow$6.33$, $6.69$\textrightarrow$6.60$). However, in the eval set, due to an unwanted increase in insertions ($1.74$\textrightarrow$1.80$), the improvement is a bit muted. The reason could be that since the RD capability of ASR (CNN-$2$) + RD (RNN) on the eval set is slightly weaker to begin with, the RD network's guidance provides only a marginal benefit.

\begin{table}[h]
    \normalsize
    \caption{Substitution, deletion, and insertion errors of ASR (CNN-$2$) + RD (RNN) model with different decoding strategies.}
    \vspace{-0.2cm}
    \centering
    \begin{adjustbox}{max width=\linewidth}
    \begin{tabular}{l|l|c|ccc}
        &ASR (CNN-$2$) + RD (RNN) &WER &SUB &DEL &INS \\
        \cline{1-6}
        \multirow{2}{*}{\rotatebox[origin=c]{90}{Val}}
        &Beam Search &14.56 &6.36 &6.47 &1.73 \\ 
        &RD-Guided Beam Search &14.29 &6.38 &6.33 &1.58 \\
        \cline{1-6}
        \multirow{2}{*}{\rotatebox[origin=c]{90}{Eval}}
        &Beam Search &15.67 &7.24 &6.69 &1.74 \\
        &RD-Guided Beam Search &15.65 &7.25 &6.60 &1.80 \\
    \end{tabular}
    \end{adjustbox}
    \label{tab:wer_sub_del_ins}
    \vspace{-0.2cm}
\end{table}

Although RD-guided blank suppression yields only marginal improvements on the evaluation set, the approach nonetheless shows clear promise. While the RD network is trained only on frames where ASR subword tokens are emitted, its activity on other frames indicates an implicit awareness of speaker activity across the entire audio. We demonstrated that this insight can help reduce small-word deletions, and future work will further investigate how to better address the suppressed-word phenomenon identified in this study.

% \textbf{Dataset path}:
% Val (800ms VAD): \text{/research/mark.fuhs/scribe/test/obsidian/val-2024-01/2024-06-release-role/vad/}\\
% Eval (800ms VAD): \text{/research/mark.fuhs/scribe/test/obsidian/eval-2024-01/2024-06-release-role/vad/}

% \textbf{Model path}:
% Main directory: \text{/research/arindam.ghosh/espnet-mmm/egs2/dopacos/asr1}
% \begin{itemize}
%     \item ASR (CNN-2): \text{exp/asr5a_ebranchformer_stateless/inference.valid.loss.ave.till40epoch}
    
%     \item ASR (RNN): \text{exp/asr4_ebranchformer_med6/inference.valid.loss.ave.till40epoch}
    
%     \item Role-ASR (CNN-2): \text{exp/role_asr5b_ebranchformer_stateless/inference.valid.loss.ave.till40epoch}
    
%     \item Role-ASR (RNN): \text{exp/role_asr4a_ebranchformer_med6/inference.valid.loss.ave.till40epoch}
    
%     \item ASR + SRD following \cite{huang24d_interspeech}: \text{exp/role_asr_blank_huang_etal}
    
%     \item ASR (CNN-2) L6 + SRD: \text{exp/role_asr_blank_6/inference.valid.wder.ave.till20epoch}
    
%     \item ASR (CNN-2) L6 + SRD (CNN-2): \text{exp/role_asr_blank_4/inference.valid.wder.ave.till30epoch}
    
%     \item ASR (CNN-2) L6 + SRD (RNN): \text{exp/role_asr_blank_5/inference.valid.wder.ave.till30epoch}
    
%     \item ASR (CNN-2) L12 + SRD (RNN): \text{exp/role_asr_blank_5_l12e/inference.valid.wder.ave.till30epoch}

%     \item ASR (CNN-2) FT: \text{exp/asr5a_ebranchformer_stateless_simeco/inference.valid.loss.ave.till40epoch}

%     \item Role-ASR (RNN) FT: \text{exp/role_asr_simeco/inference.valid.loss.ave.till30epoch}
% \end{itemize} 

\end{document}